\documentclass[aps,preprint]{revtex4-2}
\usepackage{graphicx}
\usepackage{lipsum}
\usepackage{float}
\usepackage{amsmath}
\usepackage{amssymb}
\usepackage{mathtools}
\usepackage{color}

\begin{document}

\title{
The distribution of shortest path lengths on trees of a given size in
subcritical Erd{\H o}s-R\'enyi networks
}

\author{Barak Budnick, Ofer Biham and Eytan Katzav}
\affiliation{
Racah Institute of Physics, 
The Hebrew University, 
Jerusalem 9190401, Israel}

\begin{abstract}

In the subcritical regime
Erd{\H o}s-R\'enyi (ER) networks  
consist of finite tree components,
which are non-extensive in the network size.
The distribution of shortest path lengths (DSPL) 
of subcritical ER networks was recently calculated
using a topological expansion
[E. Katzav, O. Biham and A.K. Hartmann, Phys. Rev. E 98, 012301 (2018)].
The DSPL, which accounts for the distance $\ell$ between any pair of nodes that
reside on the same finite tree component,
was found to follow a geometric
distribution of the form
$P(L=\ell | L < \infty) = (1-c) c^{\ell - 1}$,
where $0 < c < 1$ is the mean degree of the network.
This result includes the contributions of trees of all possible sizes
and topologies.
Here we calculate the distribution of shortest path lengths
$P(L=\ell | S=s)$ 
between random pairs of nodes that reside
on the same tree component of a given size $s$.
It is found that 
$P(L=\ell | S=s) = \frac{\ell+1}{s^{\ell}}   \frac{(s-2)!}{(s-\ell-1)!}$. 
Surprisingly, this distribution
does not depend on the mean degree $c$ of the network
from which the tree components were extracted.
This is due to the fact that the ensemble of tree components of a given size
$s$ in subcritical ER networks is sampled uniformly from the set of labeled trees of size $s$
and thus does not depend on $c$.
The moments of the DSPL are also calculated.
It is found that the mean distance between random pairs of nodes on
tree components of size $s$ satisfies
${\mathbb E}[L|S=s] \sim \sqrt{s}$,
unlike small-world networks in which the mean distance
scales logarithmically with $s$.

\end{abstract}


\maketitle

\newpage

\section{Introduction}

Random networks provide a useful framework for the analysis of
a large variety of systems that consist of interacting objects
\cite{Havlin2010,Estrada2011,Newman2018,Dorogovtsev2022}.
One can distinguish between two major types of random networks:
supercritical networks and subcritical networks. 
Supercritical networks form a giant component
that encompasses a macroscopic fraction of all the nodes.
The giant component may provide a useful description of networks 
in which the connectivity is essential, such as the
world-wide-web, social networks, and infrastructure networks.
The giant component is a small-world network, namely the mean
distance between pairs of nodes on the giant component scales
logarithmically with its size.
It includes a large number of cycles with a broad spectrum of cycle lengths
\cite{Marinari2004,Marinari2006,Bonneau2017}.
These cycles provide redundancy in the connectivity between pairs of nodes via multiple paths.
The redundancy helps to maintain the integrity of the giant component
upon deletion of nodes or edges due to failures or attacks.
The combination of the small-world property and the redundancy gives rise
to highly efficient channels of transport and communication and to the robustness of the network. 
In contrast, subcritical networks consist of finite tree
components that do not scale with the overall network size.
In a tree topology each pair of nodes is connected by a single path. 
Therefore, in subcritical networks the shortest path between any pair
of nodes that reside on the same tree component is, in fact, the only path between them.
As a result, in subcritical networks each node of degree $k \ge 2$ is an 
articulation point, namely its deletion would break the tree component
on which it resides into at least two disconnected parts
\cite{Tian2017,Tishby2018}.
Moreover, each edge is a bredge (bridge edge), 
namely its deletion would break the tree component
on which it resides into two disconnected parts
\cite{Bonneau2020}.
The subcritical tree components may describe
the fragmented structure of
secure compartmentalized networks,
such as the communication networks 
of commercial enterprises, government agencies
and illicit organizations
\cite{Duijn2014}.
The structure of such networks may be determined by the trade-off between efficiency and security.
When security considerations outweigh efficiency considerations, the number of communication
lines may need to be reduced to a minimum, which is achieved in the case of tree structures.
Other examples of fragmented networks include networks that suffered multiple failures, 
large scale attacks or epidemics,
in which the remaining functional or uninfected nodes form small, isolated components
\cite{Shao2008,Shao2009}.
In spite of their importance, the structural and statistical properties of subcritical networks  
have not attracted nearly as much attention as those of supercritical networks. 

Random networks of the 
Erd{\H o}s-R\'enyi (ER) type
\cite{Erdos1959,Erdos1960,Erdos1961} 
are the simplest class of random networks and
are used as a benchmark for the study of structure
and dynamics in complex networks
\cite{Bollobas2001}.
The ER network ensemble
is a maximum entropy ensemble, 
under the condition
that the mean degree
$\langle K \rangle = c$
is fixed.
It is a special case of a broader class of random uncorrelated networks,
referred to as configuration model networks
\cite{Bollobas1980,Molloy1995,Molloy1998,Newman2001}.
In an ER network of $N$ nodes, 
each pair of nodes
is independently connected with probability $p$,
such that the mean degree is
$c=(N-1)p$.
It was recently shown that the ER graph structure 
is an asymptotic structure for networks that contract due to
node deletion processes, which may result from failures, attacks or epidemics
\cite{Tishby2019,Tishby2020}.

The degree distribution of ER networks follows a
Poisson distribution of the form

\begin{equation}
P(K=k) = \frac{e^{-c} c^{k}}{k!}.
\label{eq:poisson}
\end{equation}

\noindent
ER networks exhibit a percolation transition 
at $c=1$ such that for $c>1$ (supercritical regime)
there is a giant component 
\cite{Tishby2018b},
while for $0 < c < 1$ (subcritical regime) 
the network consists of small, isolated tree components
\cite{Bollobas2001,Durrett2007}.
In the special case of $c=0$ the network consists
of $N$ isolated nodes and the degree distribution degenerates into
$P(K=k) = \delta_{k,0}$.

\begin{figure}
\begin{center}
\includegraphics[width=13cm]{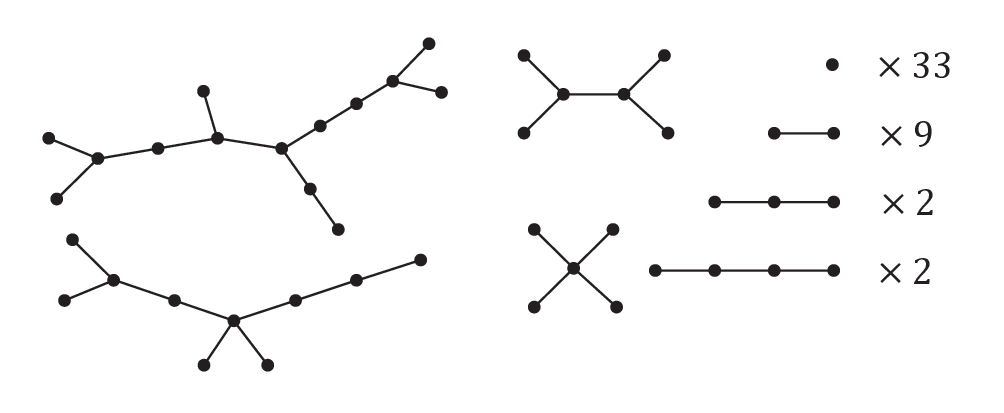}
\end{center}
\caption{
The structure of a single instance of a subcritical ER network 
of $N=100$ nodes with mean degree $c=0.9$.
It consists of 33 isolated nodes, 9 dimers, two chains of three nodes,
two chains of four nodes and trees of 5, 6, 10 and 14 nodes.
}
\label{fig:1}
\end{figure}

In Fig. \ref{fig:1} we present  
the structure of a single instance of a subcritical ER network 
of size $N=100$ with mean degree $c=0.9$.
It consists of 33 isolated nodes, 9 dimers, two chains of three nodes,
two chains of four nodes and trees of 5, 6, 10 and 14 nodes.

In the asymptotic limit, ER networks exhibit duality with
respect to the percolation threshold
\cite{Bollobas2001}. 
In a supercritical ER
network of $N$ nodes the fraction of nodes that belong to the
giant component is denoted by $0 < g \le 1$, while the fraction of nodes
that belong to the finite components is $1 - g$. Thus, the
subcritical network that consists of the finite components is of
size $N(1-g)$ . This network is in itself an ER network whose mean
degree is $c'=c(1-g)$, where $c' < 1$.

The distribution of tree sizes in subcritical ER networks with mean degree $0 < c < 1$
is given by
\cite{Bollobas2001,Newman2007,Katzav2018}

\begin{equation}
P(S=s) = \frac{ 2 s^{s-2} c^{s-1} e^{-cs} }{ (2-c)s! }.
\label{eq:PSs}
\end{equation}

\noindent
In the special case of $c=0$ this distribution degenerates into
$P(S=s)=\delta_{s,1}$.

The mean tree size is given by
\cite{Katzav2018}

\begin{equation}
\langle S \rangle = \frac{2}{2-c}.
\end{equation}

\noindent
The expected number of trees in a network instance consisting of $N$ nodes is thus given by

\begin{equation}
N_T = \frac{N}{\langle S \rangle} = N \left( 1 - \frac{c}{2} \right).
\label{eq:NT}
\end{equation}

\noindent
The variance of $P(S=s)$ is given by
\cite{Katzav2018}

\begin{equation}
{\rm Var}(S) = \frac{2c}{(1-c)(2-c)^2}.
\end{equation}

\noindent
Note that ${\rm Var}(S)$ diverges as $c \rightarrow 1^{-}$,
which implies that near the percolation transition some of the trees
are very large.

Trees of a given size $s$ may exhibit different structures, where the number of
distinct structures increases with $s$.
An important distinction in this context is between labeled trees, in which
nodes are distinguishable and carry labels, and unlabeled trees in which the nodes are indistinguishable.
The number $T_s$ of distinct labeled tree configurations of size $s$ is given by
the Cayley formula 
\cite{Cayley1889}

\begin{equation}
T_s = s^{s-2}.
\label{eq:T_s}
\end{equation}

\noindent
Each one of these labeled tree configurations can be encoded by a unique sequence,
refereed to as the Pr\"ufer sequence
\cite{Prufer1918}. 
The Pr\"ufer sequence of a labeled tree of $s$ nodes
is a string of $s-2$ integers, taking values in the range of $1,2,\dots,s$.
The Pr\"ufer code provides a very powerful tool for the random sampling of 
labeled trees of a given size.

When the labels are removed, the number of distinct configurations is reduced since each
unlabeled configuration corresponds to several labeled configurations.
In the case of unlabeled trees, the
number of non-isomorphic tree topologies, $n(s)$, which can be assembled
from $s$ nodes quickly increases as a function of $s$. 
For example, the values of $n(s)$ for $s=1,2,\dots,13$ are
1, 1, 1, 2, 3, 6, 11, 23, 47, 106, 235, 551 and 1301, respectively
\cite{Steinbach1990}.
An efficient algorithm
for generating all the tree topologies
that can be assembled from
$s$ nodes, is presented in Refs.
\cite{Wright1986,Beyer1980}.
A list of all possible tree topologies up to $s=13$ is presented in Ref.
\cite{Steinbach1990}.
 
\begin{figure}
\begin{center}
\includegraphics[width=13cm]{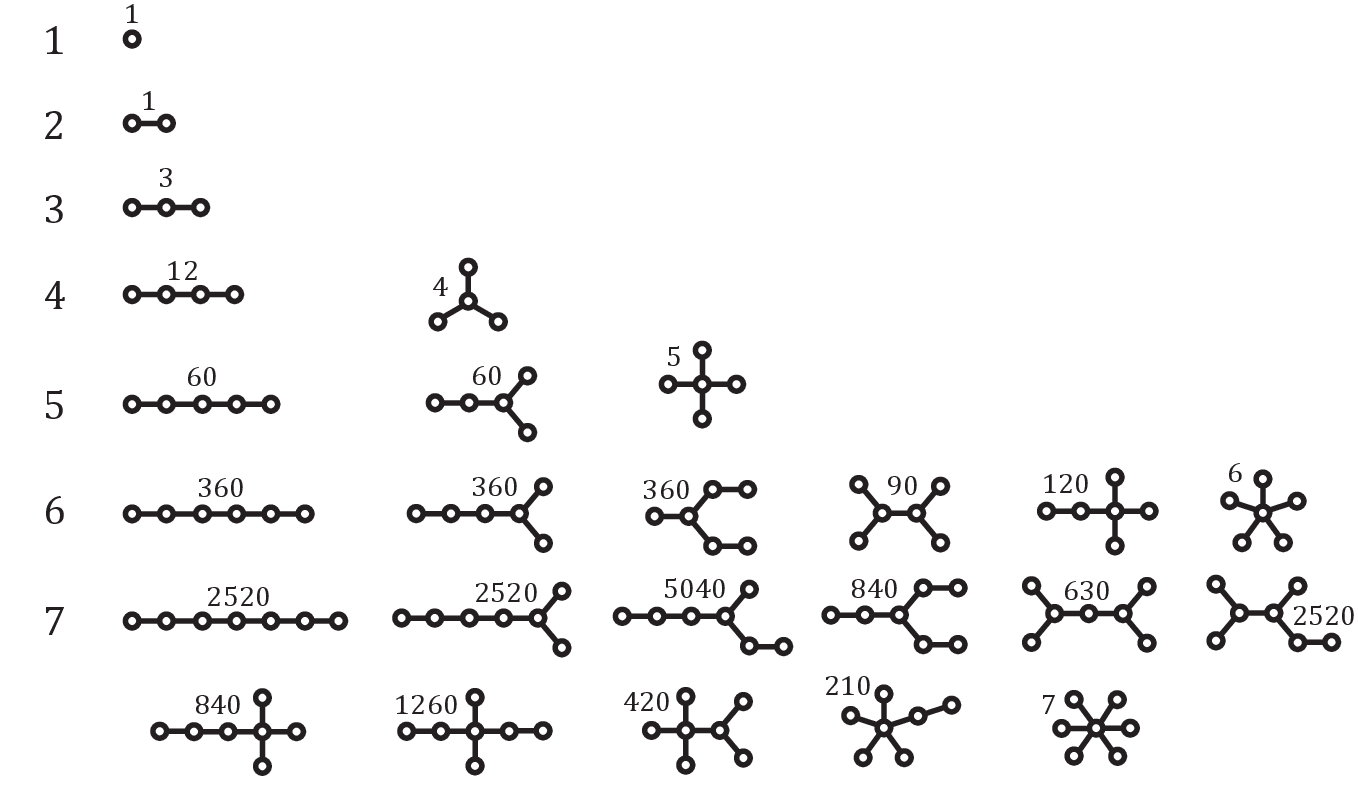}
\end{center}
\caption{
The tree topologies that consist of $s$ nodes 
for $s=1,2,\dots,7$.
For $s \le 3$ the linear chain topology is the only topology while
for $s \ge 4$ more complex topologies appear and their number
quickly increases.
The number of labeled configurations associated with each one of
the tree topologies is also shown.
Note that the total number of labeled trees that consist of $s$ nodes
add up to $s^{s-2}$, which is consistent with the Cayley formula
(\ref{eq:T_s}).
}
\label{fig:2}
\end{figure}

In Fig. \ref{fig:2} we present  
the tree topologies that consist of $s$ nodes 
for $s=1,2,\dots,7$.
For $s \le 3$ the linear chain topology is the only possible topology while
for $s \ge 4$ more complex topologies appear and their number
quickly increases.
The number of labeled configuration associated with each one of
the tree topologies is also shown.
Note that the total number of labeled trees that consist of $s$ nodes
add up to $s^{s-2}$, which is consistent with the Cayley formula
(\ref{eq:T_s}).

While the local structure of a network is well characterized by the
degree distribution, the distribution of shortest path lengths (DSPL),
denoted by $P(L=\ell)$, provides
a useful characterization of its large scale structure.
When two nodes,
$i$ and $j$, 
reside on the same connected
component, 
the distance, 
$\ell_{ij}$, 
between them
is given by the length of the shortest path 
that connects them.
When nodes $i$ and $j$ reside on different network components,
there is no path connecting them
and the distance between them is
$\ell_{ij} = \infty$.
The probability that two randomly selected
nodes reside on the same component, and
thus are at a finite distance from each other,
is denoted by
$P(L<\infty) = 1 - P(L=\infty)$. 
The conditional DSPL between pairs 
of nodes that reside on the same component is
denoted by
$P(L=\ell | L<\infty)$,
where
$\ell=1,2,\dots,N-1$.
The conditional DSPL satisfies

\begin{equation}
P(L=\ell | L<\infty)=\frac{P(L=\ell)}{P(L<\infty)}.
\end{equation}

\noindent
Note that $P(L=\ell | L<\infty)$ is well defined only for $c>0$.
This is due to the fact that $P(L<\infty)=0$ for $c=0$.
Thus, the analysis presented below is focused on $0 < c < 1$.

The DSPL provides
a natural platform for 
the study of dynamical processes on networks,
such as 
diffusive processes,
epidemic spreading,
critical phenomena,
synchronization,
information propagation and communication.
For supercritical networks 
the DSPL was calculated using various theoretical approaches,
which include recursion equations, generating functions, master equations
and branching processes
\cite{Newman2001,Dorogotsev2003,Hofstad2005,Blondel2007,Hofstad2007,Hofstad2008,Esker2008,Shao2008,Shao2009,Katzav2015,Nitzan2016,Melnik2016,Bonneau2017,Steinbock2017,Tishby2018b,Steinbock2019,Tishby2022,Jackson2022}.
In the special case of random regular graphs with $c \ge 3$ the giant component encompasses the
whole network. 
In this case there is a closed-form analytical expression
for $P(L=\ell)$ 
\cite{Hofstad2005,Nitzan2016,Tishby2022}, 
which follows a discrete Gompertz distribution
\cite{Gompertz1825}.

It was shown that the mean distance 
${\mathbb E}[ L | L < \infty]  = \sum_{\ell=1}^{\infty} \ell P(L=\ell | L<\infty)$ 
scales like
${\mathbb E}[ L | L < \infty]  \sim \ln N / \ln c$,
in agreement with rigorous results,
showing that supercritical random networks are small-world networks
\cite{Chung2002,Chung2003,Fronczak2004,Bollobas2007}.
It was also shown that the variance of the DSPL of supercritical random networks does not scale with $N$,
and satisfies ${\rm Var}(L) \sim {\mathcal O}(1)$
\cite{Nitzan2016}.
The statistical properties of distances in scale-free networks,
which typically consist of a single connected component,
were studied in Refs. 
\cite{Cohen2003,Hofstad2007,Hofstad2008}. 
Using an analytical argument it was shown that scale free networks with degree distributions of
the form 
$P(k) \sim k^{-\gamma}$
are ultrasmall, namely they exhibit a mean distance which scales like
${\mathbb E}[L] \sim \ln \ln N$ for $2 < \gamma < 3$.
For $\gamma=3$ it was shown that the mean distance scales
like ${\mathbb E}[L] \sim \ln N/ \ln \ln N$, while for $\gamma > 3$ 
it coincides with the common scaling of small
world networks, namely ${\mathbb E}[L] \sim \ln N$.

The DSPL of subcritical ER networks was recently studied using a topological expansion
\cite{Katzav2018}.
This analysis employs the fact that in the subcritical regime,
in the large-network limit,
the network consists of
finite tree components with no cycles
\cite{Bollobas2001,Durrett2007}.
It was found that for $0 < c < 1$ the
DSPL between pairs of nodes that reside on the same tree component
is given by
\cite{Katzav2018}

\begin{equation}
P(L=\ell | L<\infty ) = (1-c) c^{\ell-1},
\label{eq:PLellc}
\end{equation}

\noindent
and that the
probability that two random nodes 
reside on the same tree 
component is  
\cite{Katzav2018}

\begin{equation}
P(L < \infty) = \frac{c}{(1-c)N}.
\end{equation}

\noindent
The corresponding tail distribution is given by

\begin{equation}
P(L>\ell | L<\infty ) = c^{\ell}.
\label{eq:PLellctail}
\end{equation}

\noindent
The mean
distance between pairs of 
nodes that reside on the same tree component is  

\begin{equation}
{\mathbb E}[ L | L < \infty] = \frac{1}{1-c},
\label{eq:ELLfinite}
\end{equation}

\noindent
while the variance of the DSPL is given by

\begin{equation}
{\rm Var}(L|L<\infty) = \frac{c}{(1-c)^2}.
\label{eq:VarLLfinite}
\end{equation}

While subcritical ER networks consist of finite tree components, in
supercritical ER networks there is a coexistence between the giant
component and the finite tree components. 
As a result, the DSPL of supercritical ER networks combines the
contributions of the giant and finite components.
Using the duality relations discussed above, 
the DSPL of the finite components of a supercritical ER network can be
obtained from the analysis of its dual subcritical network 
\cite{Katzav2018,Tishby2018b}.

In this paper  
we calculate the DSPL 
of finite tree components of size $s$, 
denoted by $P(L=\ell|S=s)$,
in subcritical ER networks.
This is done by expressing the overall distribution 
$P(L=\ell)$ as a linear combination of the corresponding
conditional distributions 
$P(L=\ell|S=s)$, using
the known distribution of tree sizes.
Using an inverse transformation we extract the conditional distribution
$P(L=\ell|S=s)$.
Surprisingly, this distribution
does not depend on the mean degree $c$ of the network
from which the tree components were extracted.
This is due to the fact that the ensemble of tree components of a given size
$s$ in subcritical ER networks is sampled uniformly from the set of labeled trees of size $s$
and thus does not depend on $c$.
This insight is corroborated by a direct combinatorial argument.
We also calculate the DSPL over all tree components up to size $s$, denoted by $P(L=\ell | S \le s)$
and examine its convergence towards 
the DSPL of the whole network, $P(L=\ell | L < \infty)$, as $s$ is increased.
The moments of the DSPL are also calculated.
It is found that the mean distance between random pairs of nodes on
tree components of size $s$ satisfies
${\mathbb E}[L|S=s] \sim \sqrt{s}$,
unlike small-world networks in which the mean distance
scales logarithmically with $s$.

The paper is organized as follows.
In Sec. II we consider the conditional DSPL on finite tree components.
The moments of the DSPL are calculated in Sec. III.
The results are discussed in Sec. IV and summarized in Sec. V.

\section{The distribution of shortest path lengths}

Using the law of total probability the DSPL of subcritical ER networks,
given by Eq. (\ref{eq:PLellc}), 
can be expressed in the form

\begin{equation}
P(L=\ell | L < \infty) = \sum_{s=2}^{\infty} 
P(L=\ell| S=s) \widehat P(S=s),
\label{eq:TotalProb}
\end{equation}

\noindent
where $P(L=\ell| S=s)$ is the DSPL on tree components that consist
of $s$ nodes and $\widehat P(S=s)$ is the distribution
of tree sizes on which a pair of random nodes resides
(given that they reside on the same tree component).
In the analysis below we extract a closed-form expression for
$P(L=\ell | S=s)$ by inverting the infinite system of linear equations, 
given by Eq. (\ref{eq:TotalProb}).
Unlike commonly used methods for the calculation of such distributions, which
are based on combinatorial considerations, this approach is purely algebraic.
It is essentially a top-down approach, in which the conditional distribution 
$P(L = \ell | S=s)$ is obtained from the overall distribution 
$P(L=\ell | L<\infty)$ via the distribution of tree sizes $P(S=s)$.
This approach is advantageous over the complementary bottom-up
approach, which would require a detailed knowledge of all the tree
configurations of size $s$, their weights and the DSPL over each and
every one of them.

The distribution $\widehat P(S=s)$ is given by

\begin{equation}
\widehat P(S=s) = \frac{ \binom{s}{2} }{ \left\langle \binom{S}{2} \right\rangle } P(S=s),
\label{eq:whPSs}
\end{equation}

\noindent
where 

\begin{equation}
\left\langle \binom{S}{2} \right\rangle = \sum_{s=2}^{\infty}
\binom{s}{2} P(S=s) 
\label{eq:s_2}
\end{equation}

\noindent
is the mean number of pairs of nodes in a randomly selected tree component,
and $P(S=s)$ is given by Eq. (\ref{eq:PSs}).
This is due to the fact that the number of pairs of nodes on a tree component
of size $s$ is given by the binomial coefficient $\binom{s}{2}$.
The evaluation of $\left\langle \binom{S}{2} \right\rangle$ 
is presented in Appendix A.
It yields
 
\begin{equation}
\left\langle \binom{S}{2} \right\rangle = 
\frac{c}{(1-c)(2-c)},
\label{eq:s_2b}
\end{equation}

\noindent
where $0 < c < 1$.
Inserting $P(S=s)$ from Eq. (\ref{eq:PSs}) and 
$\langle \binom{s}{2} \rangle$ from Eq. (\ref{eq:s_2b})
into Eq. (\ref{eq:whPSs}), we obtain

\begin{equation}
\widehat P(S=s) =  
(1-c) \frac{ s^{s-2} c^{s-2} e^{-cs} }{(s-2)!}.
\label{eq:whPSs2}
\end{equation}

\noindent
Inserting 
$\widehat P(S=s)$ from Eq. (\ref{eq:whPSs2}) 
and $P(L=\ell | L < \infty)$ from Eq. (\ref{eq:PLellc})
into Eq. (\ref{eq:TotalProb}),
we obtain

\begin{equation}
\sum_{s=2}^{\infty}
\frac{ s^{s-2} c^{s-1} e^{-cs} }{ (s-2)! }
P(L=\ell|S=s) = c^{\ell}.
\end{equation}

\noindent
This equation can  be re-written in the form

\begin{equation}
\sum_{s=2}^{\infty}
\frac{ s^{s-2} (c e^{-c})^s }{ (s-2)! }
P(L=\ell|S=s) = c^{\ell+1}.
\label{eq:invert1}
\end{equation}

\noindent
The distribution $P(L=\ell | S=s)$ is obtained by inverting Eq. (\ref{eq:invert1}).
In the inversion process we assume that $P(L=\ell | S=s)$ does not depend on
the mean degree $c$. The results presented below show that such a solution indeed
exists and is justified by a combinatorial argument. The resulting expression
for $P(L=\ell | S=s)$ is verified by computer simulations.

Defining 

\begin{equation}
x = c e^{-c}
\label{eq:x_c}
\end{equation}

\noindent
enables us to express the left hand side of
Eq. (\ref{eq:invert1}) as a power series in $x$.
For the analysis below, it will be useful to also express the right hand side
in terms of $x$ rather than $c$. 
To this end, we invert Eq. (\ref{eq:x_c}) and obtain

\begin{equation}
c = -W(-x),
\label{eq:c_x}
\end{equation}

\noindent
where $W(x)$ is the Lambert $W$ function
\cite{Olver2010}.
Eq. (\ref{eq:invert1}) can now be written in the form

\begin{equation}
\sum_{s=2}^{\infty}
\frac{ s^{s-2}  }{ (s-2)! }
P(L=\ell|S=s) 
x^s
= 
[ - W(-x) ]^{\ell+1}.
\label{eq:sumPL}
\end{equation}

\noindent
From equation (3.2.2) in Ref. \cite{Gessel2016},
which results from the Lagrange inversion formula,
we obtain the identity

\begin{equation}
[ W(x) ]^r = (-r) \sum_{s=r}^{\infty} 
\frac{ (-s)^{s-r-1} }{ (s-r)! }
x^s.
\label{eq:Identity}
\end{equation}

\noindent
Using Eq. (\ref{eq:Identity}) we now express the right hand side of Eq. (\ref{eq:sumPL}) 
as a power series in $x$.
Comparing the coefficients of $x^s$ on both sides of Eq. (\ref{eq:sumPL}),
we obtain the DSPL of tree components that consist of $s$ nodes 
in subcritical ER networks with $0 < c < 1$.
It is given by

\begin{equation}
P(L=\ell|S=s) = 
\frac{ (\ell+1) }{s^{\ell}}
\frac{ (s-2)! }{ (s-\ell-1)! },
\label{eq:PLs}
\end{equation}

\noindent
where $s \ge 2$ and $1 \le \ell \le s-1$.
This is the central result of the paper.
Clearly, this distribution does not depend on the mean degree $c$ of the subcritical 
network from which the trees of size $s$ were extracted.

Unlike the DSPL of the whole network, which is a monotonically decreasing geometric
distribution, $P(L=\ell | S=s)$ exhibits a peak.
The location of the peak is referred to as the mode of the distribution
and is denoted by $\ell_{\rm mode}$.
Since $P(L=\ell | S=s)$ exhibits a single peak, $\ell_{\rm mode}$ is the
lowest integer for which $P(L=\ell+1 | S=s) < P(L=\ell | S=s)$.
Using Eq. (\ref{eq:PLs}), this inequality can be expressed in the form

\begin{equation}
\left( \frac{\ell +2}{\ell+1} \right) \left( \frac{s-\ell-1}{s} \right) < 1.
\end{equation}

\noindent
The solution of this inequality (assuming positive $\ell$) is 

\begin{equation}
\ell > \frac{ \sqrt{4s+1} - 3}{2}.
\label{eq:ellgt}
\end{equation}

\noindent
The mode $\ell_{\rm mode}$ is the lowest integer that satisfies
Eq. (\ref{eq:ellgt}), namely

\begin{equation}
\ell_{\rm mode} = \left\lceil \frac{ \sqrt{4 s + 1} - 3 }{2} \right\rceil,
\label{eq:mode}
\end{equation}

\noindent
where $\lceil x \rceil$ is the lowest integer that is larger than $x$,
also known as the ceiling function.
In the limit of large trees, the mode scales like 
$\ell_{\rm mode} \sim \sqrt{s}$.

It turns out that the DSPL given by Eq. (\ref{eq:PLs}) 
coincides with the DSPL of the ensemble obtained by uniformly random
sampling over all the labeled tree configurations of size $s$
\cite{Meir1970,Moon1970}.
The DSPL over all the labeled tree configurations of size $s$
can be obtained from
direct combinatorial considerations.
To this end we pick a random pair of nodes $i$ and $j$ 
on a tree of size $s$.
We count the number of possible configurations of labeled trees of size $s$,
in which the distance between a given pair of nodes $i$ and $j$ is $\ell$.
The fact that the distance between $i$ and $j$ is $\ell$ 
implies that there is a single path of length $\ell$ between them.
This path consists of $\ell-1$ intermediate nodes.
The number of ways to select these $\ell-1$ nodes from the $s-2$ nodes
(not including $i$ and $j$), where the order is important, is given by

\begin{equation}
\frac{ (s-2)! }{ (s-\ell-1)! } = \binom{s-2}{\ell-1} (\ell-1)!.
\end{equation}

\noindent
The path joining $i$ and $j$, which consists of $\ell+1$ nodes
(including $i$ and $j$),
can be considered as the backbone of the tree.
Each node on the backbone may be the root of a tree branch such that
each one of the remaining $s-\ell-1$ nodes belongs to one of these tree branches.
This enables us to use the generalized Cayley formula
\cite{Cayley1889,Takacs1990,Shor1995}, 
which provides the
number of labeled tree configurations that consist of $\ell+1$ non-empty disjoint tree components
(also known as forests)
with a total of $s$ nodes,
namely

\begin{equation}
T_{s,\ell+1} = (\ell+1) s^{s-\ell-2}.
\label{eq:T_sell}
\end{equation}

\noindent
Note that Cayley formula of Eq. (\ref{eq:T_s}) is a special case of the
generalized Cayley formula (\ref{eq:T_sell}), namely
$T_s = T_{s,1}$.
The probability $P(L=\ell | S=s)$ is obtained by dividing
the number of possible configurations of labeled trees of size $s$,
in which the distance between a given pair of nodes $i$ and $j$ is $\ell$
by the total number $T_s$ of configurations of labeled trees of size $s$.
It yields

\begin{equation}
P(L=\ell | S=s) = \frac{ T_{s,\ell+1} \binom{s-2}{\ell-1} (\ell-1)! }{T_{s}},
\label{eq:PLs2}
\end{equation}

\noindent
which is equivalent to Eq. (\ref{eq:PLs}).
This equivalence suggests that the ensemble of trees of a given size $s$ in subcritical ER networks
is equivalent to a uniformly random sampling among all the $T_s$ labeled tree configurations
of size $s$.
This is consistent with the fact that the DSPL given by Eqs. (\ref{eq:PLs})
and (\ref{eq:PLs2})
does not depend on the mean degree $c$ of the network from which these 
trees were extracted.
The equivalence between the two ensembles
can be justified using the following argument.
Given a finite connected component consisting of $s$ nodes in a subcritical ER network
it is almost surely to exhibit a tree topology containing $s-1$ edges
\cite{Bollobas2001}.
For a set of $s$ nodes, the
probability that these nodes will form a connected tree component
of a given labeled configuration,
which is isolated from the rest of the network, 
is given by

\begin{equation}
p^{s-1} (1-p)^{ \binom{s}{2} - (s-1) } (1-p)^{s (N-s)},
\end{equation}

\noindent
where the first term accounts for the $s-1$ edges of the tree,
the second term accounts for the probability that there are no
additional edges between the nodes in the tree component,
and the third term accounts
for the probability that the tree is isolated from the rest of the network.
In an ER network, in which the connectivity between different pairs of nodes
is independent, this probability is the same for all possible configurations of labeled trees
of size $s$.

Summing up the right hand side of Eq. (\ref{eq:PLs}) from $\ell+1$ to infinity,
we obtain the tail distribution, which is given by

\begin{equation}
P(L>\ell | S=s) = \frac{(s-2)!}{s^{s-2}} \frac{s^{s - \ell - 2}}{(s - \ell - 2)!},
\end{equation}

\noindent
where $\ell = 0,1,2,\dots,s-2$.
It is a monotonically decreasing function that satisfies
$P(L>0 | S=s) = 1$ and
$P(L>s-2 | S=s) = (s-2)!/s^{s-2}$.

\begin{figure}
\begin{center}
\includegraphics[width=7cm]{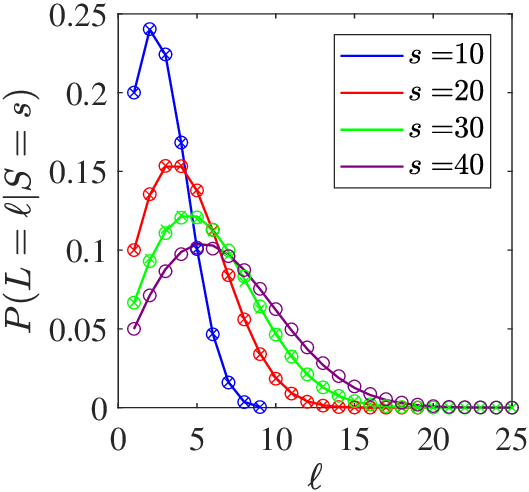}
\end{center}
\caption{
(Color online)
analytical results (solid lines) for the DSPL on trees of size $s$,
denoted by $P(L=\ell | S=s)$, for $s=10$, $20$, $30$ and $40$ 
(left to right),
obtained from Eq. (\ref{eq:PLs}).
The analytical results are in very good agreement with the results obtained 
from computer simulations carried out for networks of size $N=10^4$,
$c=0.5$ ($\times$) and
$c=0.8$ ($\circ$), which coincide with each other.
These results confirm the validity of Eq. (\ref{eq:PLs}) as well as the fact that the ensemble of
finite trees of a given size $s$ extracted from subcritical ER networks of mean degree
$c$ does not depend on $c$.
Note that the simulation results for $c=0.5$ are shown only for $s=10$, $20$ and $30$,
because trees of size $s=40$ are extremely rare in this case.
}
\label{fig:3}
\end{figure}

In Fig. \ref{fig:3} we present analytical results (solid lines) for the DSPL on trees of size $s$,
denoted by $P(L=\ell | S=s)$, for $s=10$, $20$, $30$ and $40$,
obtained from Eq. (\ref{eq:PLs}).
The analytical results are in very good agreement with the results obtained 
from computer simulations carried out for 
$c=0.5$ ($\times$) and
$c=0.8$ ($\circ$), which coincide with each other.
These results confirm the validity of Eq. (\ref{eq:PLs}) as well as the fact that the ensemble of
finite trees of a given size $s$ extracted from subcritical ER networks of mean degree
$c$ does not depend on $c$.

In the simulations we generated subcritical ER networks of size $N=10^4$
with mean degree $c=0.5$ and $c=0.8$. 
From these networks we picked tree components of the desired sizes,
such as $s=10$, $20$, $30$ and $40$.
The expected number of trees of size $s$ in a network instance of size $N$
is given by 

\begin{equation}
N_T(s) = N_T P(S=s).
\label{eq:NTs}
\end{equation}

\noindent
Inserting $N_T$ from Eq. (\ref{eq:NT}) and $P(S=s)$ from Eq. (\ref{eq:PSs})
into Eq. (\ref{eq:NTs}), we obtain

\begin{equation}
N_T(s) = N \frac{ s^{s-2} c^{s-1} e^{-cs} }{s!}.
\label{eq:NTs2}
\end{equation}

\noindent
This result can be used in order to estimate the number of network instances which
is required in order to obtain the desired number of trees of size $s$ that are needed for
the statistical analysis.
The distribution $P(S=s)$ is a quickly decreasing function of $s$.
Thus, trees of size $s$ become less abundant as $s$ is increased.
As a result, one needs a large number of network instances in order to
obtain sufficient data for statistical analysis of large tree components.
The results presented in Fig. \ref{fig:3} are based on 1,500 instances of subcritical ER
networks of size $N=10^4$ for each value of $c$.
For $c=0.8$ these network instances yield 12,454  trees of size 10,
1,823 trees of size 20, 500 trees of size 30 and 183 trees of size 40.
For $c=0.5$ these network instances yield 3,617 trees of size 10,
91 trees of size 20, 8 trees of size 30 and no trees of size 40.
Therefore, In Fig. \ref{fig:3} the analytical results for $s=40$ are compared only to
the simulation results for $c=0.8$ ($\circ$).

Another interesting distribution is the
DSPL between pairs of nodes that reside on all tree components of size
$s' \le s$. It can be obtained from 

\begin{equation}
P(L=\ell|S \le s) =
\frac{ \sum_{s'=2}^s 
\widehat P(S=s') P(L=\ell|S=s') }
{\sum_{s'=2}^s 
\widehat P(S=s')}.
\label{eq:PLSscums}
\end{equation}

\noindent
Taking the limit of large $s$, $P(L=\ell|S \le s)$ converges towards 
$P(L=\ell | L < \infty)$, as in Eq. (\ref{eq:TotalProb}). 
To explore this convergence it is convenient to replace the
sums $\sum_{s'=2}^s$ in Eq. (\ref{eq:PLSscums}) by the difference  
$\sum_{s'=2}^{\infty} - \sum_{s'=s+1}^{\infty}$. 
Carrying out the first summations in the numerator and in the
denominator, we obtain

\begin{equation}
P(L=\ell|S \le s) =
\frac{  (1-c)c^{\ell-1} -  
\sum_{s'=s+1}^{\infty} 
\widehat P(S=s') P(L=\ell|S=s') }
{1 - \sum_{s'=s+1}^{\infty} 
\widehat P(S=s')}.
\label{eq:PLSscums2}
\end{equation}

\begin{figure}
\begin{center}
\includegraphics[width=7cm]{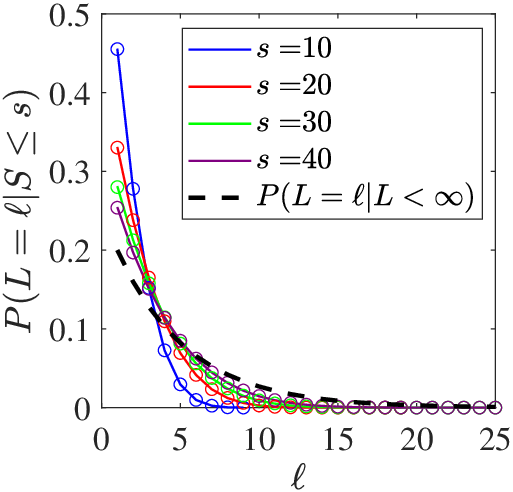}
\end{center}
\caption{
(Color online)
Analytical results (solid lines) for $P(L=\ell|S \le s)$ on tree components
of size smaller or equal to $s$ in subcritical ER network with mean degree $c=0.8$,
for $s=10$, $20$, $30$ and $40$ (top to bottom on the left hand side),
obtained from Eq. (\ref{eq:PLSscums2}).
As $s$ is increased, $P(L=\ell|S \le s)$ converges towards
the overall DSPL, $P(L=\ell | L<\infty)$, 
of the subcritical ER network (dashed line).
The analytical results are in very good agreement with the results obtained 
from computer simulations ($\circ$).
}
\label{fig:4}
\end{figure}

In Fig. \ref{fig:4} we present analytical results (solid lines) for the distribution
$P(L=\ell | S \le s)$ of shortest path lengths on all tree components of size smaller
or equal to $s$, in subcritical ER networks with mean degree $c=0.8$.
The analytical results 
obtained from Eq. (\ref{eq:PLSscums2}),
are presented for tree sizes of
$s=10$, $20$, $30$ and $40$
(top to bottom on the left hand side).
The analytical results are in very good agreement with the results obtained 
from computer simulations carried out for $c=0.8$ ($\circ$).
As $s$ is increased, the distribution $P(L=\ell | S \le s)$ converges towards
the overall DSPL 
$P(L=\ell | L<\infty)$
of the subcritical ER network (dashed line).
  
\section{The mean and variance of the DSPL}


In order to calculate the moments of the DSPL, we define
the moment generating function

\begin{equation}
M(x) = 
\sum_{\ell=1}^{s-1}
e^{x \ell} P(L=\ell|S=s).
\label{eq:MLs}
\end{equation}

\noindent
Inserting the probability $P(L=\ell|S=s)$
from Eq. (\ref{eq:PLs})
into Eq. (\ref{eq:MLs})
and carrying out the summation,
we obtain

\begin{equation}
M(x) = \frac{s}{s-1}
\left[ \left( e^{-2x} - \frac{1}{s} \right) 
+ \frac{e^{-x}}{s^s} \left( 1 - e^{-x} \right) e^{s \left( x + e^{-x} \right)}
\Gamma \left( s+1,s e^{-x} \right) \right].
\label{eq:MLs2}
\end{equation}

\noindent
where $\Gamma(a,z)$ is the incomplete Gamma function
\cite{Olver2010}.
The $n$th moment of $P(L=\ell | S=s)$ is obtained by differentiating
$M(x)$, with respect to $x$, $n$ times, namely

\begin{equation}
{\mathbb E} \left[ L^n | S=s \right] = \frac{ \partial^n M }{\partial x^n} \bigg\vert_{x=0}.
\label{eq:ELns}
\end{equation}

\noindent
Inserting $n=1$ in Eq. (\ref{eq:ELns}), we obtain the mean distance between 
random pairs of nodes that reside on a tree component of size $s$.
It is given by

\begin{equation}
{\mathbb E}[L|S=s] = 
\frac{ s[e^{s} s^{-s} \Gamma(s+1,s) - 2 ] }{s-1}.
\label{eq:ELs1}
\end{equation}

\noindent
Inserting $n=2$ in Eq. (\ref{eq:ELns}), we obtain the second moment,
which is given by

\begin{equation}
{\mathbb E}[L^2|S=s] = 
\frac{ s[4 + 2s - 3 e^{s} s^{-s} \Gamma(s+1,s) ] }{s-1}.
\label{eq:EL2s}
\end{equation}



\noindent
The variance of $P(L=\ell | S=s)$ is given by

\begin{equation}
{\rm Var}(L|S=s) = 
{\mathbb E}[L^2|S=s] 
-
\left( {\mathbb E}[L|S=s] \right)^2,
\label{eq:varLs}
\end{equation}

\noindent
where
${\mathbb E}[L^2|S=s]$
is given by Eq. (\ref{eq:EL2s})
and
${\mathbb E}[L|S=s]$
is given by Eq. (\Ref{eq:ELs1}).

For sufficiently large values of $s$ one can obtain simplified asymptotic 
expressions for the moments of the DSPL.
To achieve this we use the double-asymptotic expansion of $\Gamma(s,s)$,
given by equation 8.11.12 in Ref. \cite{Olver2010},
namely

\begin{equation}
\Gamma(s,s) = s^{s-1} e^{-s} \left[ \sqrt{ \frac{\pi}{2} } \sqrt{s} 
- \frac{1}{3} + \mathcal{O} \left( \frac{1}{\sqrt{s} } \right) \right]
\label{eq:Gamma_ss}
\end{equation}

\noindent
To evaluate the moments, we need a closed form expression for
$\Gamma(s+1,s)$.
Using equation 8.8.2 in Ref. \cite{Olver2010},
we obtain

\begin{equation}
\Gamma(s+1,s) = s \Gamma(s,s) + s^s e^{-s},
\end{equation}

\noindent
where $\Gamma(s,s)$ is given by Eq.  (\ref{eq:Gamma_ss}).
Equipped with these expressions, we can now obtain asymptotic 
expansions for the moments in the limit of large $s$.
More specifically, the mean distance on a random tree of size $s$
is given by

\begin{equation}
{\mathbb E}[L|S=s] = \sqrt{ \frac{\pi}{2} } \sqrt{s} - \frac{4}{3}
+ \mathcal{O} \left( \frac{1}{\sqrt{s} } \right).
\label{eq:ELSasymp}
\end{equation}

\noindent
It is found that the mean distance between random pairs of nodes that
reside on a tree component of size $s$ scales like square root of $s$.
Comparing the right hand sides of Eqs. (\ref{eq:mode}) and (\ref{eq:ELSasymp}), 
which show the mode $\ell_{\rm mode}$ and the mean distance ${\mathbb E}[L|S=s]$, respectively,
it is found that while both of them scale like $\sqrt{s}$ the pre-factor of the
mean distance is larger than the pre-factor of the mode.
This implies that the distribution $P(L=\ell | S=s)$ is positively skewed.
Interestingly, the scaling of the mean distance,
implied by Eq. (\ref{eq:ELSasymp}),
resembles the scaling of distances on two 
dimensional lattices.
It is in contrast with small world random networks in which the mean
distance scales like $\ln s$.
This means that the tree components in subcritical ER networks are
not small world networks.

\begin{figure}
\begin{center}
\includegraphics[width=7cm]{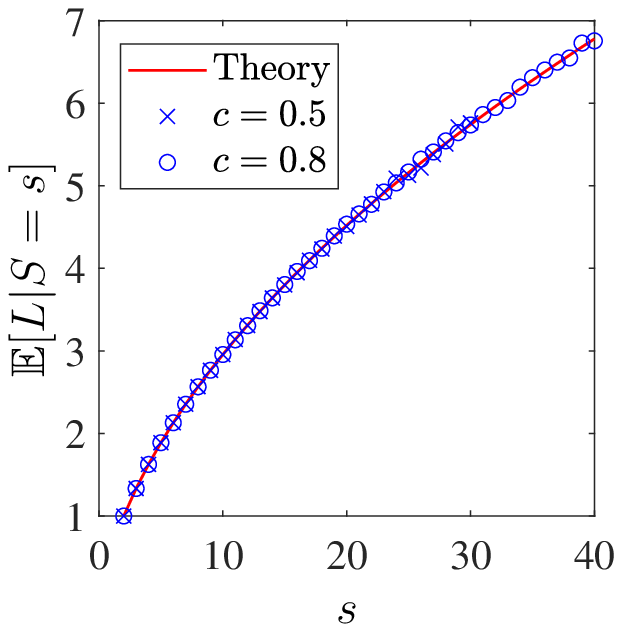}
\end{center}
\caption{
(Color online)
Analytical results (solid line) for the mean distance 
${\mathbb E}[L|S=s]$ between pairs of nodes that
reside on the same tree component of size $s$, in a subcritical ER network,
as a function of $s$. The analytical results are in very good agreement
with the results obtained from computer simulations for subcritical
ER networks of size $N=10^4$ and $c=0.5$ ($\times$) and $c=0.8$ ($\circ$),
which coincide with each other.
Note that the simulation results for $c=0.5$ are shown only up to $s=30$,
because in this case larger trees are rare.
}
\label{fig:5}
\end{figure}

In Fig. \ref{fig:5} we present 
analytical results (solid line) for the mean distance 
${\mathbb E}[L|S=s]$ between pairs of nodes that
reside on the same tree component of size $s$, in a subcritical ER network,
as a function of $s$. The analytical results are in very good agreement
with the results obtained from computer simulations for subcritical
ER networks of size $N=10^4$ with $c=0.5$ ($\times$) and $c=0.8$ ($\circ$),
which coincide with each other.
Note that the simulation results for $c=0.5$ are shown only up to $s=30$,
because in this case larger trees are rare.

The second moment of the DSPL can be expressed by

\begin{equation}
{\mathbb E}[L^2|S=s] = 2 s -
3 \sqrt{ \frac{\pi}{2} } \sqrt{s} +2   
+ \mathcal{O} \left( \frac{1}{\sqrt{s} } \right).
\end{equation}

\noindent
Combining the results presented above for the first and second moments, 
we obtain an asymptotic expression for the variance. It is given by

\begin{equation}
{\rm Var}(L|S=s) = 
\frac{ 4 - \pi }{2} s -
\sqrt{ \frac{\pi}{18} } \sqrt{s} 
+ \mathcal{O} \left(  1 \right).
\end{equation}

\noindent
Thus, the standard deviation of the DSPL on trees of size $s$ scales like
$\sqrt{s}$, namely it scales like the mean distance ${\mathbb E}[L|S=s]$.
Interestingly, the same qualitative relation is found in the DSPL of the whole
subcritical ER network.
This implies that $P(L=\ell | S=s)$ is relatively broad distribution,
in contrast with the typical results for the DSPL of supercritical 
configuration model networks
\cite{Katzav2015,Nitzan2016,Tishby2022}.

\begin{figure}
\begin{center}
\includegraphics[width=7cm]{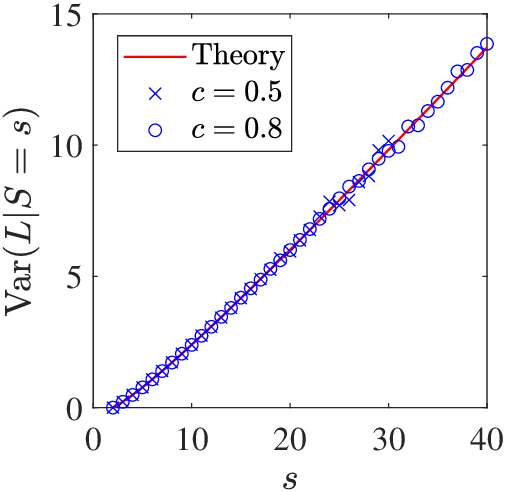}
\end{center}
\caption{
(Color online)
The variance ${\rm Var}(L|S=s)$ of the DSPL
between pairs of nodes that reside on the same tree component of size $s$, 
in a subcritical ER network,
as a function of $s$.
The analytical results are in very good agreement
with the results obtained from computer simulations
for subcritical
ER networks of size $N=10^4$ and mean degree $c=0.5$ ($\times$) and $c=0.8$ ($\circ$),
which coincide with each other.  
Note that the simulation results for $c=0.5$ are shown only up to $s=30$,
because in this case larger trees are rare.
}
\label{fig:6}
\end{figure}

In Fig. \ref{fig:6} we present analytical results (solid line) for the
variance ${\rm Var}(L|S=s)$ of the distribution of shortest path lengths
between pairs of nodes that reside on the same tree component of size $s$, 
in a subcritical ER network, as a function of $s$.
The analytical results are in very good agreement
with the results obtained from computer simulations
for subcritical
ER networks of size $N=10^4$ and mean degree $c=0.5$ ($\times$) and $c=0.8$ ($\circ$),
which coincide with each other.  
Note that the simulation results for $c=0.5$ are shown only up to $s=30$,
because in this case larger trees are rare.

The cumulative mean distance between pairs of nodes that reside on
a tree of size smaller or equal to $s$ is given by

\begin{equation}
{\mathbb E}[L|S \le s] =
\frac{ \sum_{s'=2}^s 
\widehat P(S=s') {\mathbb E}[L|S=s'] }
{\sum_{s'=2}^s 
\widehat P(S=s')}.
\label{eq:ELcums}
\end{equation}

\noindent 
To evaluate the right hand side of Eq. (\ref{eq:ELcums}),
it is convenient to express the numerator and the denominator
as differences between two infinite sums, namely

\begin{equation}
{\mathbb E}[L|S \le s] =
\frac{ \sum_{s'=2}^{\infty} 
\widehat P(S=s') {\mathbb E}[L|S=s'] 
-
\sum_{s'=s+1}^{\infty}
\widehat P(S=s') {\mathbb E}[L|S=s']
}
{\sum_{s'=2}^{\infty} 
\widehat P(S=s')
-
\sum_{s'=s+1}^{\infty}
\widehat P(S=s')
}.
\label{eq:ELcumsp}
\end{equation}

\noindent
The first term in the numerator amounts to 
${\mathbb E}[L | L < \infty]$,
which is given by Eq. (\ref{eq:ELLfinite}),
while the first term in the denominator is equal to $1$
(due to the normalization of $\widehat P(S=s)$).
Eq. (\ref{eq:ELcumsp}) can thus be simplified to

\begin{equation}
{\mathbb E}[L|S \le s] =
\left( \frac{1}{1-c} \right)
\frac{ 1 
-
(1-c) \sum_{s'=s+1}^{\infty}
\widehat P(S=s') {\mathbb E}[L|S=s']
}
{1
-
\sum_{s'=s+1}^{\infty}
\widehat P(S=s')
}.
\label{eq:ELcumspp}
\end{equation}

\noindent
Inserting $\widehat P(S=s)$ from Eq. (\ref{eq:whPSs2}) and ${\mathbb E}[L|S=s]$
from Eq. (\ref{eq:ELSasymp}), which is accurate for sufficiently large $s$,
into Eq. (\ref{eq:ELcumspp}) and carrying out the summations,
we obtain

\begin{equation}
{\mathbb E}[L|S \le s] \simeq
\left( \frac{1}{1-c} \right)
\frac{   
1 - \frac{(1-c)^2}{\sqrt{2 \pi} c^2} (c e^{1-c})^{s+1}
\left[  \sqrt{ \frac{\pi}{2} } \frac{1}{1-c e^{1-c}}  -
\frac{4}{3}
\Phi \left( c e^{1-c}, \frac{1}{2}, s+1 \right) 
\right]
}
{ 1 - \frac{1-c}{\sqrt{2 \pi} c^2} (c e^{1-c})^{s+1}
\left[ \Phi \left( c e^{1-c}, \frac{1}{2}, s+1 \right) -
\Phi \left( c e^{1-c}, \frac{3}{2}, s+1 \right) 
\right] },
\label{eq:ELcums2}
\end{equation}

\noindent
where 

\begin{equation}
\Phi(z,s,a) = \sum_{n=0}^{\infty} \frac{ z^n }{ (a+n)^s }
\label{eq:phi}
\end{equation}

\noindent
is the Lerch Phi transcendent
\cite{Olver2010}.
Eq. (\ref{eq:ELcums2}) is expected to be valid for large values of $s$.

\begin{figure}
\begin{center}
\includegraphics[width=7cm]{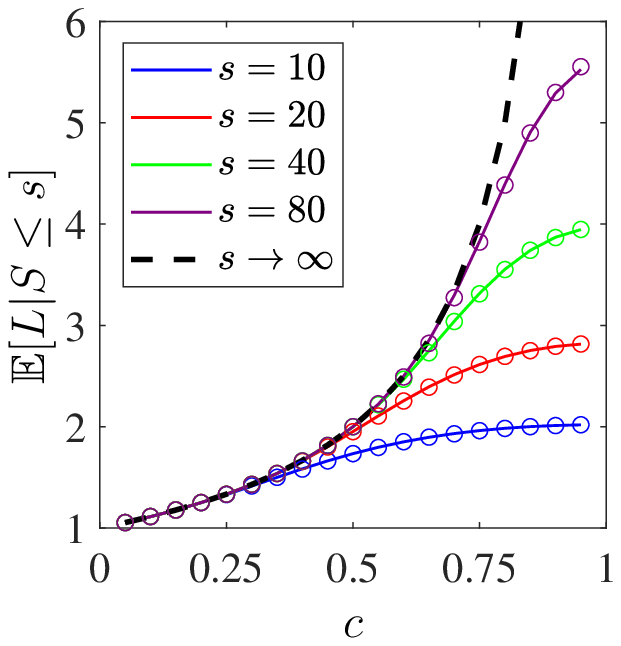}
\end{center}
\caption{
(Color online)
Analytical results (solid lines), obtained from Eq. (\ref{eq:ELcums2}), 
for the mean distance ${\mathbb E}[L|S \le s]$ 
between all pairs of nodes that
reside on the same tree component, for all tree components of size smaller or
equal to $s$, in subcritical networks, as a function of the mean degree $c$.
The results are presented for $s=10$, $20$, $40$ and $80$ (from bottom to top).
The analytical results are in very good agreement with the results obtained 
from computer simulations ($\circ$).
As $s$ is increased, the mean distance ${\mathbb E}[L|S \le s]$ converges towards
the mean distance over the whole network, ${\mathbb E}[L|L<\infty]$ (dashed line), 
given by Eq. (\ref{eq:ELLfinite}).
}
\label{fig:7}
\end{figure}

In Fig. \ref{fig:7} we present analytical results (solid lines)  
for the mean distance ${\mathbb E}[L|S \le s]$ 
between pairs of nodes that
reside on the same tree component, for all tree components of size smaller or
equal to $s$, in subcritical networks, as a function of the mean degree $c$.
The results are presented for $s=10$, $20$, $40$ and $80$ (from bottom to top).
The analytical results,
obtained from Eq. (\ref{eq:ELcums2}), 
are in very good agreement with the results obtained 
from computer simulations ($\circ$).
As $s$ is increased, the mean distance ${\mathbb E}[L|S \le s]$ converges towards
the mean distance over the whole network, ${\mathbb E}[L|L<\infty]$ (dashed line), 
given by Eq. (\ref{eq:ELLfinite}).

\section{Discussion}

The ensemble of trees that appear in subcritical ER networks belong
to the class of equilibrium trees
\cite{Dorogovtsev2022}.
These are trees that are formed by equilibrium processes. Their statistical
properties can be analyzed using methods of equilibrium statistical mechanics.
In this paper we calculated the DSPL 
of trees of a given size $s$ in subcritical ER networks.
It was found that $P(L=\ell | S=s)$ is independent of the mean 
degree $c$ of the subcritical network from which these trees were extracted.
It was also found that the mean distance on the ensemble of trees
of size $s$ scales like
${\mathbb E}[L|S=s] \sim \sqrt{s}$.
This scaling implies that the Hausdorff dimension of the trees is
$D_H=2$, in agreement with earlier results obtained 
for other equilibrium trees
\cite{Dorogovtsev2022}.
It is in contrast with the scaling obtained in supercritical ER networks
and other configuration model networks. In these networks
the mean distance
${\mathbb E}[L]$ scales logarithmically with the network size $N$
and they are thus referred to as small-world networks.

Another important ensemble of trees consists of random recursive
trees, which belong to the class of nonequilibrium trees.
These trees grow via a kinetic process of node addition.
The simplest model of random tree growth is the random attachment model.
In this model,
starting from a small seed network, at each time step a new node is
added and is connected to one of the existing nodes uniformly at random.
For simplicity we consider the case in which the seed network consists
of a single node.
Interestingly, the ensembles of equilibrium and nonequilibrium trees
of size $s$ include the same set of tree configurations.
However, their statistical properties are different due to the different
weights assigned to each one of the possible configurations.
In growing trees the order in which the nodes are added
is important.
In particular, nodes that appeared early in the growth process are
likely to gain more links than nodes that appeared at later stages
\cite{Dorogovtsev2022}.

The DSPL of the ensemble of random attachment trees of size $s$
was found to follow a Poisson distribution whose mean is
given by ${\mathbb E}[L|S=s] = 2 \ln s$ 
\cite{Steinbock2017}.
This implies that the random attachment trees belong to the
class of small-world networks, in which the mean distance
scales logarithmically with the network size.
These trees tend to form compact structures dominated
by the nodes that appeared early in the growth process.
This is in sharp contrast to the results obtained for the
subcritical ER trees in which the mean distance scales
like $\sqrt{s}$.

The methodology presented in this paper can be applied to 
the calculation of the distribution $P(L=\ell | S=s)$ in
configuration model networks with various degree distributions $P(K=k)$,
such as the exponential distribution and the power-law distribution.
To this end, one needs to obtain the distribution $P(S=s)$ of tree sizes
in the subcritical configuration model network under study and the
DSPL of the whole network, $P(L=\ell | L<\infty)$ and to insert them into 
Eq. (\ref{eq:TotalProb}).
The distribution $P(S=s)$ can be calculated using the generating function
approach presented in Ref. \cite{Newman2007}.
The inversion of Eq. (\ref{eq:TotalProb}) to extract $P(L=\ell | S=s)$ is possible probably
in those cases in which $P(L=\ell | S=s)$ is independent of the mean degree $c$.
The validity of this condition will need to be tested on a case-by-case basis.

Apart from the DSPL there are other metric properties that characterize the large
scale structure of finite trees in subcritical configuration model networks.
These include the distributions of eccentricities and diameters of trees of size $s$.
The eccentricity is a property of a single node $i$ and it is equal to the largest distance
between the given node $i$ and any other node in the tree.
The diameter is a property of the whole tree and it is equal to the largest
distance between any pair of nodes in the tree.
The distribution of the largest diameter among all the trees in a subcritical
ER network was recently studied
\cite{Luczak1998,Hartmann2018}.
It was found that this distribution follows a Gumbel distribution
\cite{Gumbel1935}, which is one of the three
distributions encountered in extreme-value theory.

The resistance distance between two nodes in a network is a measure of 
how difficult it is for electricity (or some other form of flow) to pass between 
these two nodes. In an unweighted network, the resistance distance is defined as the resistance 
between the two nodes, where the resistance of 
each edge is equal to $1$ Ohm. The resistance distance can 
be thought of as a generalization of the concept of distance to networks, 
where the "distance" between two nodes is determined by the flow resistance 
between them rather than their physical separation. 
A more formal definition is given in 
\cite{Deza2016,Bapat2014}, 
where it is also shown that it is a proper metric, 
satisfying for example the triangle inequality.
In general, the resistance distance between two nodes will be smaller 
if there are more paths between the two nodes with lower resistance, 
and larger if there are fewer paths or if the paths have higher resistance. 
Random networks of resistors have been studied,
mainly in two dimensions 
\cite{Derrida1982}, 
and recently calculated for supercritical ER networks 
\cite{Akara2022,Akara2022b}.
Interestingly, on tree graphs the shortest path between 
a pair of nodes $i$ and $j$ is in fact the only path between them. 
As a result, the resistance distance between $i$ and $j$ is equal to 
the shortest path length between them. 
This means that the results presented in this paper provide also the distribution of resistance
distances in ER networks in the subcritical regime.

\section{Summary}

We calculated the distribution of shortest path lengths
$P(L=\ell | S=s)$ 
between random pairs of nodes that reside
on finite tree components of a given size $s$
in subcritical ER networks.
It was found that 
$P(L=\ell | S=s) = \frac{\ell+1}{s^{\ell}} \frac{(s-2)!}{(s-\ell-1)!}$. 
Surprisingly, this probability
does not depend on the mean degree $c$ of the network
from which these tree components were extracted.
This is due to the fact that the ensemble of tree components of a given size
$s$ in ER networks is sampled uniformly from the set of labeled trees of size $s$.
The moments of the DSPL were also calculated.
It was found that the mean distance between random pairs of nodes on
tree components of size $s$ satisfies
${\mathbb E}[L|S=s] \sim \sqrt{s}$,
unlike small-world networks in which the mean distance
scales logarithmically with $s$. 

This work was supported by the Israel Science Foundation grant no. 
1682/18.

\appendix

\section{The generating function of $P(S=s)$}

The generating function of $P(S=s)$ is given by

\begin{equation}
H(u) = \sum_{s=1}^{\infty} u^s P(S=s).
\label{eq:Gsu}
\end{equation}

\noindent
Inserting $P(S=s)$ from Eq. (\ref{eq:PSs}) into Eq. (\ref{eq:Gsu}), we obtain

\begin{equation}
H(u) = \frac{2}{2-c} \sum_{s=1}^{\infty}
\frac{ s^{s-2} c^{s-1} e^{-cs} }{ s! } u^s.
\label{eq:Gsu2}
\end{equation}

\noindent
Rearranging terms on the right hand side of Eq. (\ref{eq:Gsu2}), we obtain

\begin{equation}
H(u) = - \frac{2}{c(2-c)} \sum_{s=1}^{\infty} 
\frac{1}{s} \frac{ (-s)^{s-1} }{ s! } \left( - u c e^{-c} \right)^s.
\label{eq:Gsu3}
\end{equation}

\noindent
Replacing the term $1/s$ on the right hand side of Eq. (\ref{eq:Gsu3}) by the 
integral expression

\begin{equation}
\frac{1}{s} = \int_{0}^{\infty} e^{ - s \tau } d \tau,
\end{equation}

\noindent
yields

\begin{equation}
H(u) = - \frac{2}{c(2-c)} \sum_{s=1}^{\infty} \int_{0}^{\infty} e^{ - s \tau } d \tau 
\frac{ (-s)^{s-1} }{ s! } \left( - u c e^{-c} \right)^s.
\label{eq:Gsu4}
\end{equation}

\noindent
Exchanging the order of the sum and the integral on the right hand side of 
Eq. (\ref{eq:Gsu4}), we obtain 

\begin{equation}
H(u) = - \frac{2}{c(2-c)} \int_{0}^{\infty} d \tau      \sum_{s=1}^{\infty}    
\frac{ (-s)^{s-1} }{ s! } \left( - u c e^{-c} e^{ - \tau } \right)^s.
\label{eq:Gsu4p}
\end{equation}

\noindent
Using the series expansion of the Lambert W function, which is given by

\begin{equation}
W(x) = \sum_{s=1}^{\infty} \frac{ (-s)^{s-1} x^s }{s!},
\end{equation}

\noindent
we obtain

\begin{equation}
H(u) = - \frac{2}{c(2-c)} \int_{0}^{\infty} d \tau W \left(- u c e^{-c} e^{- \tau} \right).
\label{eq:Gsu5}
\end{equation}

\noindent
Changing the integration variable from $\tau$ to $x = - u c e^{-c} e^{-\tau}$,
we obtain

\begin{equation}
H(u) =   \frac{2}{c(2-c)} \int_{ - u c e^{-c} }^{0}   W \left( x \right) \frac{dx}{x}.
\label{eq:Gsu6}
\end{equation}

\noindent
Changing the integration variable again, from $x$ to $y=W(x)$,
which from the definition of the Lambert function
implies that $x = y e^{y}$, we obtain

\begin{equation}
H(u) =   \frac{2}{c(2-c)} \int_{ W \left(- u c e^{-c} \right) }^{0}   y 
\left( 1 + \frac{1}{y} \right)  dy.
\label{eq:Gsu7}
\end{equation}

\noindent
Carrying out the integration on the right hand side of Eq. (\ref{eq:Gsu7}), we obtain

\begin{equation}
H(u) = - \frac{1}{c(2-c)} \left\{ \left[ W( - u c e^{-c} ) \right]^2 + 2 W ( - u c e^{-c} ) \right\}.
\end{equation}

\noindent
The moments of $P(S=s)$ can be obtained by taking suitable derivatives of $H(u)$.
In particular, the mean tree size is

\begin{equation}
\langle S \rangle = \frac{ d H(u) }{du} \bigg\vert_{u=1} = \frac{2}{2-c}.
\end{equation}

\noindent
and the second factorial moment is given by

\begin{equation}
\langle S(S-1) \rangle = \frac{ d^2 H(u) }{du^2} \bigg\vert_{u=1} = \frac{2c}{(1-c)(2-c)}.
\end{equation}

\noindent
Using these results, it is found that the second moment of $P(S=s)$ is given by

\begin{equation}
\langle S^2 \rangle =  \frac{2}{(1-c)(2-c)},
\end{equation}

\noindent
and the variance is given by

\begin{equation}
{\rm Var}(S) = \frac{2c}{(1-c)(2-c)^2}.
\end{equation}

\noindent
It is also found that

\begin{equation}
\left\langle \binom{S}{2} \right\rangle = \frac{c}{(1-c)(2-c)}.
\end{equation}



\end{document}